\documentclass[twocolumn,notitlepage,nofootinbib]{revtex4-2}
\usepackage{mathtools,amssymb,bm,graphicx,hyperref}
\allowdisplaybreaks[1]

\newcommand\grad{\bm{\nabla}}
\newcommand\<{\langle}
\renewcommand\>{\rangle}
\renewcommand\l{{\ell}}
\newcommand\R{{\bm{R}}}
\renewcommand\r{{\bm{r}}}
\newcommand\x{{\bm{x}}}
\newcommand\z{{\bm{z}}}
\newcommand\N{\mathbb{N}}
\newcommand\Z{\mathbb{Z}}
\newcommand\E{\mathcal{E}}
\newcommand\eV{\mathrm{eV}}
\newcommand\Eh{E_\mathrm{h}}
\renewcommand\H{\mathrm{H}}
\DeclareMathOperator\sgn{sgn}

\begin{document}

\title{Efimovian states of three charged particles}

\author{Yusuke Nishida}
\affiliation{Department of Physics, Tokyo Institute of Technology,
Ookayama, Meguro, Tokyo 152-8551, Japan}

\date{November 2021}

\begin{abstract}
When three particles in three dimensions interact with a short-range potential fine-tuned to an infinite scattering length, they form an infinite sequence of loosely bound states obeying discrete scale invariance known as Efimov states.
Here we show that analogous states are formed by three charged particles carrying two equal charges and one opposite charge in one, two, and three dimensions without any fine-tuning.
Our finding is based on the Born-Oppenheimer approximation, where an effective inverse-square attraction is induced as a consequence of the dipole-charge interaction between a hydrogenlike heavy-light atom and a far-separated heavy particle.
Because the resulting Efimovian states emerge toward the second or higher dissociation threshold, they are to be realized as quasibound states and may be observed by exciting hydrogen molecular ions and trions in excitonic systems.
We also consider the same system but with a logarithmic Coulomb potential relevant to quantum vortices in two-dimensional superfluids, where the Efimovian states are shown to emerge as genuine bound states toward the first dissociation threshold.
\end{abstract}

\maketitle
%\tableofcontents

\section{Introduction}
When particles interact with a short-range potential at a large scattering length, their low-energy physics becomes universal, i.e., independent of details of the short-range potential~\cite{Braaten:2006}.
The most remarkable phenomenon is the Efimov effect, predicting that three particles in three dimensions form an infinite sequence of loosely bound states obeying discrete scale invariance~\cite{Efimov:1970,Efimov:1973}.
Although the Efimov effect was theoretically discovered in the context of nuclear physics, it was experimentally observed with ultracold atomic gases~\cite{Kraemer:2006} and helium atoms~\cite{Kunitski:2015}.
Because of its universality, the Efimov effect has also been studied in diverse systems such as nucleons~\cite{Braaten:2003}, pions~\cite{Hyodo:2014}, halo nuclei~\cite{Fedorov:1994}, magnons~\cite{Nishida:2013}, and even at the Kardar-Parisi-Zhang roughening transition~\cite{Nakayama:2021}.

The Efimov effect is understood most transparently based on the Born-Oppenheimer approximation assuming that two particles are much heavier than the other particle~\cite{Fonseca:1979}.
When the light particle is bounded by the two heavy particles, its binding energy serves as an effective interaction between the two heavy particles.
With a short-range potential between the heavy and light particles fine-tuned to an infinite scattering length, the resulting effective interaction at a large separation $R$ compared to the potential range must be a scale-invariant attraction of $1/R^2$, which leads to an infinite sequence of loosely bound states obeying discrete scale invariance~\cite{Landau-Lifshitz}.
Because short-range potentials are essential to the low-energy universality, long-range potentials such as Coulomb are usually regarded as obstacles to the Efimov effect~\cite{Hammer:2008}.

What we show in this paper is that an infinite sequence of states analogous to the Efimov states is actually formed by three charged particles carrying two equal charges and one opposite charge in one, two, and three dimensions without any fine-tuning.
This is accomplished not only for a three-dimensional Coulomb potential depending inversely on an interparticle separation (Sec.~\ref{sec:charge}) but also for a two-dimensional Coulomb potential depending logarithmically on an interparticle separation (Sec.~\ref{sec:vortex}).
Although our derivations of such ``Efimovian states'' based on the Born-Oppenheimer approximation are rather elementary, they shall be described in a self-contained manner so as to make the underlying physics transparent.

\section{Three-dimensional Coulomb potential}\label{sec:charge}
\subsection{Born-Oppenheimer approximation}
Let us study two heavy particles with masses $M_1$ and $M_2$ and charge $+q$ and one light particle with $m$ and $-q$ in $d=1$, 2, and 3 dimensions, which are described by
\begin{align}
& E\Phi(\R_1,\R_2,\r_3)
= \left(-\sum_{i=1,2}\frac{\hbar^2\nabla_{\!\R_i}^2}{2M_i}
- \frac{\hbar^2\nabla_{\!\r_3}^2}{2m}\right. \notag\\
&\left.{} + \frac{k_eq^2}{|\R_2-\R_1|}
- \sum_{i=1,2}\frac{k_eq^2}{|\r_3-\R_i|}\right)\Phi(\R_1,\R_2,\r_3).
\end{align}
Here $k_e=1/4\pi\epsilon_0$ is the Coulomb constant and the Born-Oppenheimer approximation for a large mass ratio $M_1,M_2\gg m$ factorizes the wave function as
\begin{align}
\Phi(\R_1,\R_2,\r_3) = \psi_{\R_{1,2}}(\r_3)\Psi(\R_1,\R_2)
\end{align}
and neglects $\grad_{\!\R_1}$ and $\grad_{\!\R_2}$ acting on $\psi_{\R_{1,2}}(\r_3)$.
Consequently, the light particle adjusts its wave function according to
\begin{align}\label{eq:charge_light}
\E_{\R_{1,2}}\psi_{\R_{1,2}}(\r_3)
= \left(-\frac{\hbar^2\nabla_{\!\r_3}^2}{2m}
- \sum_{i=1,2}\frac{k_eq^2}{|\r_3-\R_i|}\right)\psi_{\R_{1,2}}(\r_3)
\end{align}
for given positions of the heavy particles, whereas the heavy particles slowly move according to
\begin{align}\label{eq:charge_heavy}
E\Psi(\R_1,\R_2)
&= \left(-\sum_{i=1,2}\frac{\hbar^2\nabla_{\!\R_i}^2}{2M_i}
+ \frac{k_eq^2}{|\R_2-\R_1|} + \E_{\R_{1,2}}\right) \notag\\
&\quad \times \Psi(\R_1,\R_2)
\end{align}
under an effective interaction of $\E_{\R_{1,2}}$ induced by the light particle~\cite{Landau-Lifshitz}.

\subsection{Linear Stark effect}
\begin{figure}[t]
\includegraphics[width=0.9\columnwidth]{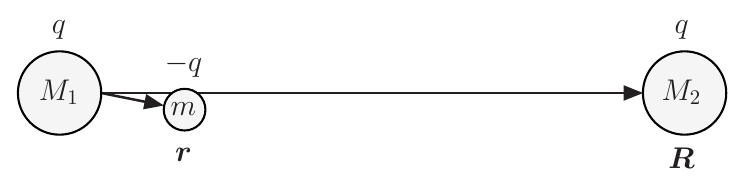}
\caption{\label{fig:stark}
Schematic configuration where heavy ($M_1$) and light ($m$) particles form a hydrogenlike atom and its binding energy is lowered by a far-separated heavy particle ($M_2$) due to the linear Stark effect.
The resulting energy shift scales as $-1/R^2$, leading to the dipole-charge interaction.}
\end{figure}

When the two heavy particles are far separated, $|\R_2-\R_1|\to\infty$, the light particle is localized around one of them, $\r_3\sim\R_1$, so as to form a hydrogenlike atom.
The Schr\"odinger equation~(\ref{eq:charge_light}) for the light particle in this limit is reduced to
\begin{align}
\E_\R\psi_\R(\r) &= \left[-\frac{\hbar^2\nabla_{\!\r}^2}{2m}
- \frac{k_eq^2}{r} - \frac{k_eq^2}{R}\right. \notag\\
&\quad\left.{} - \frac{k_eq^2}{R^2}\hat\R\cdot\r + O(R^{-3})\right]\psi_\R(\r)
\end{align}
with $\R\equiv\R_2-\R_1$ and $\r\equiv\r_3-\R_1$, where the light particle is subjected to a uniform electric field produced by the far-separated heavy particle (see Fig.~\ref{fig:stark}).
The binding energy of the light particle up to $O(R^{-2})$ corrections can be obtained with the first-order perturbation theory by regarding $(k_eq^2/R^2)\hat\R\cdot\r$ as a small perturbation, which is none other than the linear Stark effect for a hydrogenlike atom~\cite{Landau-Lifshitz}.
We introduce the Bohr radius via $a_0\equiv\hbar^2/(mk_eq^2)$, and the bound-state solutions to the hydrogenlike problems in all $d=1$, 2, and 3 dimensions are reviewed in the Appendix.

The nondegenerate ground state does not exhibit the linear Stark effect because its wave function is isotropic.
The first excited states for $d=3$ are fourfold degenerate and spanned by $|n,\l,m_\l\>=|2,0,0\>$, $|2,1,0\>$, and $|2,1,\pm1\>$, where $n$ refers to the principal quantum number, and $\l$ and $m_\l$ refer to the angular momenta.
By choosing $\hat\R=\hat\z$, the perturbation term is diagonalized on the basis of
\begin{subequations}\begin{align}
\frac{|2,0,0\>\pm|2,1,0\>}{\sqrt2} \quad\Rightarrow\quad \<\hat\R\cdot\r\> &= \mp3a_0, \\
|2,1,\pm1\> \quad\Rightarrow\quad \<\hat\R\cdot\r\> &= 0.
\end{align}\end{subequations}
Similarly, the first excited states for $d=2$ are threefold degenerate and spanned by $|n,\l\>=|2,0\>$ and $|2,\pm1\>$, where $\l$ refers to the angular momentum~\cite{Yang:1991}.
By choosing $\hat\R=\hat\x$, the perturbation term is diagonalized on the basis of
\begin{subequations}\begin{align}\hspace{-6mm}
\frac{\sqrt2\,|2,0\>\pm(|2,+1\>+|2,-1\>)}{2}
\quad\Rightarrow\quad \<\hat\R\cdot\r\> &= \mp\frac{9a_0}{4}, \\
\frac{|2,+1\>-|2,-1\>}{\sqrt2} \quad\Rightarrow\quad \<\hat\R\cdot\r\> &= 0.
\end{align}\end{subequations}
Finally, the first excited states for $d=1$ are twofold degenerate and spanned by $|n,\l\>=|2,0\>$ and $|2,1\>$, where $\l=0$ and 1 refer to even and odd parity, respectively~\cite{Loudon:1959}.
By choosing $\hat\R=\hat\x$, the perturbation term is diagonalized on the basis of
\begin{align}
\frac{|2,0\>\pm|2,1\>}{2} \quad\Rightarrow\quad \<\hat\R\cdot\r\> = \mp\frac{3a_0}{2}.
\end{align}
Although the same analysis can be carried out for every higher excited state~\cite{Landau-Lifshitz}, it is not be pursued here.

Therefore, the lowest-energy state at $n=2$ in each dimension has $\<\hat\R\cdot\r\>=3(d+1)a_0/4$, corresponding to the light particle mostly on the side of the far-separated heavy particle (see Fig.~\ref{fig:stark}), and its binding energy is found to be
\begin{align}\label{eq:charge_energy}
\E_\R = \E_{n=2} - \frac{k_eq^2}{R} - \frac{3(d+1)\hbar^2}{4mR^2} + O(R^{-3}).
\end{align}
Here the first term on the right-hand side is the unperturbed excited-state energy of a hydrogenlike atom presented in Eq.~(\ref{eq:hydrogen}), whereas the rest originate from the Coulomb potential produced by the far-separated heavy particle.
In particular, the third term is the energy shift due to the linear Stark effect.

\subsection{Efimovian states}\label{sec:efimov}
With Eq.~(\ref{eq:charge_energy}) substituted into the Schr\"odinger equation~(\ref{eq:charge_heavy}) for the heavy particles, we obtain
\begin{align}
E\Psi(\R) = \left[-\frac{\hbar^2\nabla_{\!\R}^2}{2M}
+ \E_{n=2} - \frac{3(d+1)\hbar^2}{4mR^2} + O(R^{-3})\right]\Psi(\R),
\end{align}
where the center-of-mass motion is separated and $M\equiv M_1M_2/(M_1+M_2)$ is the reduced mass.
We note that the Coulomb potentials $\sim1/R$ cancel out and the residual effective interaction induced by the light particle is dominated by the scale-invariant attraction of $1/R^2$ at a large separation $R\gg a_0$.
This is none other than the dipole-charge interaction with the dipole always pointing to the charge as a consequence of fast motion of the light particle.

It is now straightforward to show that the two heavy particles form an infinite sequence of loosely bound states obeying discrete scale invariance~\cite{Landau-Lifshitz}.
By separating the radial and angular variables as in Eq.~(\ref{eq:separation}) with the angular momentum (parity for $d=1$) denoted by $L$, the radial wave function for $E<\E_{n=2}$ is provided by the modified Bessel function in the form of $\Psi_L(R)=R^{1-d/2}K_{is_L}(\kappa R)$, where $\kappa\equiv\sqrt{2M(\E_{n=2}-E)/\hbar^2}$ and
\begin{align}\label{eq:scaling}
s_L = \sqrt{\frac{3(d+1)M}{2m} - \left(L+\frac{d-2}{2}\right)^2}.
\end{align}
Because of $\Psi_L(R)\to R^{1-d/2}|\Gamma(is_L)|\cos[s_L\ln(\kappa R/2)-\arg\Gamma(is_L)]$ for $\kappa\to0$, any boundary condition imposed on $\Psi_L(R)$ at $R\sim a_0$ can be satisfied by an infinite sequence of $\kappa\sim a_0^{-1}e^{-\pi N/s_L}$ ($N\in\Z$), so that the binding energies are found to be
\begin{align}
E_N = \E_{n=2} - \frac{\hbar^2\kappa_L^{*2}}{2M}e^{-2\pi N/s_L} \quad (N\gg1).
\end{align}
Here the scaling exponent $s_L$ depends on the dimensionality, the mass ratio, and the angular momentum, whereas the prefactor $\kappa_L^*\sim a_0^{-1}$ defined up to multiplicative factors of $e^{\pi/s_L}$ can be determined by computing the binding energies with the full effective interaction for an arbitrary $R$~\cite{Bates:1968,Madsen:1971}.

Each sequence emergent for $L$ satisfying $|L+(d-2)/2|<\sqrt{3(d+1)M/2m}$ is twofold degenerate except for possible degeneracies due to magnetic and spin quantum numbers.
This is because the light atom can be localized around either $\R_1$ or $\R_2$ and the exchange energy splitting between gerade and ungerade orbitals is exponentially small.
On the other hand, when the two heavy particles are identical bosons or fermions with $M_1=M_2$, each sequence becomes nondegenerate because only gerade or ungerade orbital is allowed depending on the parity of $L$.
We also note that all the results presented so far hold even in the case where the one heavy particle at $\R_2$ has the opposite charge of $-q$ provided that $\<\hat\R\cdot\r\>=-3(d+1)a_0/4$ is chosen.
In this case, each sequence is nondegenerate.

The resulting infinite sequence of loosely bound states obeying discrete scale invariance constitutes our Efimovian states of three charged particles.
It should be remarked that they actually emerge above the first dissociation threshold at $\E_{n=1}$ corresponding to the hydrogenlike atom in its ground state and the unbound heavy particle.
Therefore, the Efimovian states beyond the Born-Oppenheimer approximation are to be realized as quasibound states embedded in the continuum, which are similar to four-body Efimov states~\cite{Hammer:2007,Stecher:2009} and atomic collapse states~\cite{Shytov:2007,Wang:2013}.
Because the Born-Oppenheimer approximation is supposed to be valid for a sufficiently large mass ratio, we expect that the Efimovian states have small widths and are thus observable as sharp resonances.
In fact, the width to binding energy ratio of Efimov states was found to be exponentially small as $\Gamma_N/E_N\sim e^{-\#\sqrt{M/m}}$, as well as being independent of $N$ so as to keep the discrete scale invariance intact~\cite{Pen'kov:1999}.

\section{Two-dimensional Coulomb potential}\label{sec:vortex}
\subsection{Born-Oppenheimer approximation}
It is known that charged particles with a logarithmic Coulomb potential are realized by quantum vortices in two-dimensional superfluids~\cite{Chaikin-Lubensky}, three of which carrying two equal charges and one opposite charge are described by
\begin{align}
& E\Phi(\R_1,\R_2,\r_3)
= \left[-\sum_{i=1,2}\frac{\hbar^2\nabla_{\!\R_i}^2}{2M_i}
- \frac{\hbar^2\nabla_{\!\r_3}^2}{2m}\right. \notag\\
&\left.{} - KQ^2\ln\!\left(\frac{|\R_2-\R_1|}{\delta}\right)
+ \sum_{i=1,2}KQ^2\ln\!\left(\frac{|\r_3-\R_i|}{\delta}\right)\right] \notag\\
&\qquad \times \Phi(\R_1,\R_2,\r_3).
\end{align}
Here the effective Coulomb constant $K$ and the charge $Q$ for $d=2$ correspond to the mass density of a superfluid and the circulation of a quantum vortex, respectively, whereas all $d=1$, 2, and 3 dimensions shall be considered for generality.
$\delta$ is an arbitrary length scale and irrelevant to physics because it only provides a constant energy shift.
Therefore, we set $\delta=b_0$ with $b_0\equiv\sqrt{\hbar^2/(mKQ^2)}$ being the effective Bohr radius, which is equivalent to shifting the total energy as $E\to E+KQ^2\ln(b_0/\delta)$.

Again, within the Born-Oppenheimer approximation for a large mass ratio $M_1,M_2\gg m$, the above Schr\"odinger equation for three charged particles is separated into those for the one light particle,
\begin{align}\label{eq:vortex_light}
\E_{\R_{1,2}}\psi_{\R_{1,2}}(\r_3)
&= \left[-\frac{\hbar^2\nabla_{\!\r_3}^2}{2m}
+ \sum_{i=1,2}KQ^2\ln\!\left(\frac{|\r_3-\R_i|}{b_0}\right)\right] \notag\\
&\quad \times \psi_{\R_{1,2}}(\r_3),
\end{align}
and for the two heavy particles,
\begin{align}\label{eq:vortex_heavy}
E\Psi(\R_1,\R_2)
&= \Biggl[-\sum_{i=1,2}\frac{\hbar^2\nabla_{\!\R_i}^2}{2M_i}
- KQ^2\ln\!\left(\frac{|\R_2-\R_1|}{b_0}\right) \notag\\
&\quad + \E_{\R_{1,2}}\Biggr]\,\Psi(\R_1,\R_2),
\end{align}
where $\E_{\R_{1,2}}$ serves as an effective interaction induced by the light particle.
 
\subsection{Quadratic Stark effect}
When the two heavy particles are far separated, $|\R_2-\R_1|\to\infty$, the light particle is localized around one of them, $\r_3\sim\R_1$, so as to form a heavy-light atom.
The Schr\"odinger equation~(\ref{eq:vortex_light}) for the light particle in this limit is reduced to
\begin{align}
\E_\R\psi_\R(\r) = \left[H + KQ^2\ln\!\left(\frac{R}{b_0}\right)
+ V + O(R^{-3})\right]\psi_\R(\r),
\end{align}
where the unperturbed Hamiltonian is
\begin{align}
H \equiv -\frac{\hbar^2\nabla_{\!\r}^2}{2m} + KQ^2\ln\!\left(\frac{r}{b_0}\right)
\end{align}
and
\begin{align}
V \equiv - KQ^2\frac{\hat\R\cdot\r}{R} + KQ^2\frac{r^2-2(\hat\R\cdot\r)^2}{2R^2}
\end{align}
is regarded as a small perturbation.
The first term in $V$ is a uniform electric field produced by the far-separated heavy particle, and the binding energy of the light particle up to $O(R^{-2})$ corrections can be obtained with the second-order perturbation theory.

\begin{figure}[t]
\includegraphics[width=0.9\columnwidth]{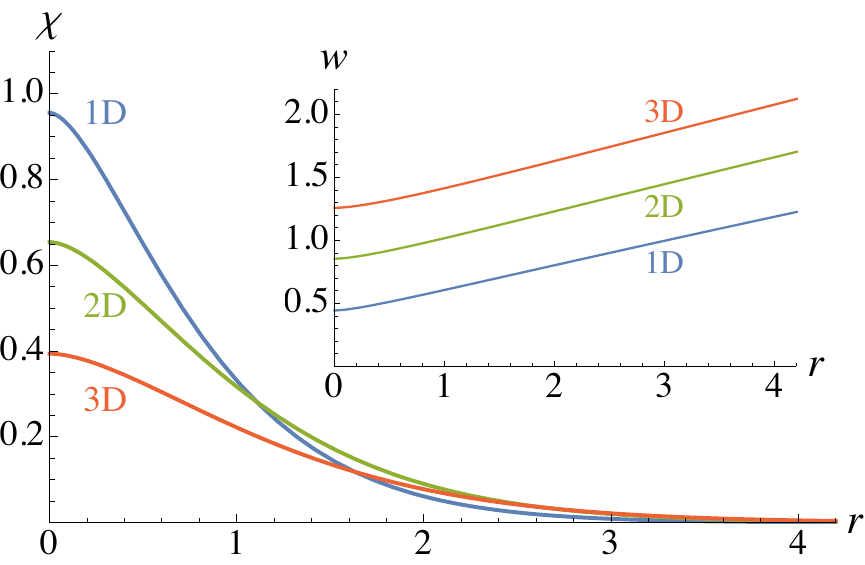}
\caption{\label{fig:chi-w_r}
$\chi(r)$ and $w(r)$ (inset) as functions of $r$ in units of $KQ^2=b_0=1$ for $d=3$ (red line), $d=2$ (green line), and $d=1$ (blue line), where $\chi(r)$ is normalized as $\int\!d\r[\chi(r)]^2=1$.}
\end{figure}

The ground state of the unperturbed Hamiltonian is determined by solving $H\chi(r)=\E\chi(r)$ for $\ell=0$, where the ground-state energy is numerically found to be
\begin{align}\label{eq:unperturbed}
\E_{n=1} = KQ^2 \times
\begin{dcases}
\,0.6978 & (d=3), \\
\,0.1799 & (d=2), \\
-0.8764 & (d=1),
\end{dcases}
\end{align}
and the corresponding wave function is plotted in Fig.~\ref{fig:chi-w_r}.
The first-order correction to the ground-state energy is then provided by
\begin{align}\label{eq:1st-order}
\E'_{n=1} = \<V\> = \frac{KQ^2}{R^2}\frac{d-2}{2d}\<r^2\>
\end{align}
with
\begin{align}
\<r^2\> = b_0^2 \times
\begin{dcases}
\,2.399 & (d=3), \\
\,1.091 & (d=2), \\
\,0.2858 & (d=1).
\end{dcases}
\end{align}
On the other hand, the second-order correction to the ground-state energy is none other than the quadratic Stark effect and reads
\begin{align}
\E''_{n=1} &= -\sum_{n\neq1}\frac{\<\chi|V|n\>\<n|V|\chi\>}{\E_n-\E_{n=1}} \\
&= -\frac{(KQ^2)^2}{R^2}\<(\hat\R\cdot\r)W(\r)\> + O(R^{-3}),
\end{align}
where an auxiliary function of coordinates is introduced via $[H,W(\r)]|\chi\>=\hat\R\cdot\r|\chi\>$~\cite{Landau-Lifshitz,Dalgarno:1955}.
The resulting differential equation for $W(\r)$,
\begin{align}
-\frac{\hbar^2}{2m}\nabla^2W(\r)
-\frac{\hbar^2}{m}\frac{\chi'(r)}{\chi(r)}\,\hat\r\cdot\grad W(\r)
= \hat\R\cdot\r,
\end{align}
is solved numerically by substituting $W(\r)=(\hat\R\cdot\r)w(r)$, whose solution plotted in Fig.~\ref{fig:chi-w_r} leads to $\<(\hat\R\cdot\r)W(\r)\>=\<r^2w(r)\>/d$ with
\begin{align}
\<r^2w(r)\> = \frac{b_0^2}{KQ^2} \times
\begin{dcases}
\,3.935 & (d=3), \\
\,1.230 & (d=2), \\
\,0.1720 & (d=1).
\end{dcases}
\end{align}

Finally, with all the results put together, the binding energy of the light particle is found to be
\begin{align}\label{eq:vortex_energy}
\E_\R &= \E_{n=1} + KQ^2\ln\!\left(\frac{R}{b_0}\right)
- \frac{\hbar^2}{mR^2} \times
\begin{dcases}
\,0.9117 & (d=3) \\
\,0.6152 & (d=2) \\
\,0.3149 & (d=1)
\end{dcases} \notag\\
&\quad + O(R^{-3}).
\end{align}
Here the first term on the right-hand side is the unperturbed ground-state energy of a heavy-light atom in Eq.~(\ref{eq:unperturbed}), whereas the rest originate from the logarithmic Coulomb potential produced by the far-separated heavy particle.
In particular, the third term is the energy shift solely due to the quadratic Stark effect for $d=2$ because of the vanishing first-order correction in Eq.~(\ref{eq:1st-order}), leading to the induced dipole-charge interaction.

\subsection{Efimovian states}
With Eq.~(\ref{eq:vortex_energy}) substituted into the Schr\"odinger equation~(\ref{eq:vortex_heavy}) for the heavy particles, we obtain
\begin{align}
E\Psi(\R) = \left[-\frac{\hbar^2\nabla_{\!\R}^2}{2M}
+ \E_{n=1} - \frac{\hbar^2C_d}{mR^2} + O(R^{-3})\right]\Psi(\R),
\end{align}
where the center-of-mass motion is separated and $C_d=0.3149$, 0.6152, and 0.9117 for $d=1$, 2, and 3, respectively, are the numerical constants.
We note that the logarithmic Coulomb potentials $\sim\ln R$ cancel out and the residual effective interaction induced by the light particle is dominated by the scale-invariant attraction of $1/R^2$ at a large separation $R\gg b_0$.
Consequently, as described in Sec.~\ref{sec:efimov}, the two heavy particles form an infinite sequence of loosely bound states for $L$ satisfying $|L+(d-2)/2|<\sqrt{2C_dM/m}$.
Their binding energies are provided by
\begin{align}
E_N = \E_{n=1} - \frac{\hbar^2\kappa_L^{*2}}{2M}e^{-2\pi N/s_L} \quad (N\gg1),
\end{align}
obeying discrete scale invariance under the scaling exponent of
\begin{align}
s_L = \sqrt{\frac{2C_dM}{m} - \left(L+\frac{d-2}{2}\right)^2}.
\end{align}
We note that all the remarks in Sec.~\ref{sec:efimov} regarding the degeneracy of each sequence also apply here.
More importantly, the Efimovian states resulting from the logarithmic Coulomb potential prove to be realized as genuine bound states emergent below the first dissociation threshold at $\E_{n=1}$.

\section{Summary and prospects}
In summary, we showed that three charged particles carrying two equal charges and one opposite charge form an infinite sequence of quasibound states obeying discrete scale invariance in all dimensions without any fine-tuning.
Our finding of such Efimovian states is based on the Born-Oppenheimer approximation assuming that two particles are much heavier than the other particle, which is potentially relevant to diverse systems in atomic and molecular physics, condensed matter physics, and nuclear and hadron physics.
Promising candidates include trions, i.e, bound states of an electron-hole pair with another electron or hole in excitonic systems~\cite{Kezerashvili:2019}, not to mention hydrogen molecular ions~\cite{Roth:2008}.

\begin{figure}[t]
\includegraphics[width=0.9\columnwidth]{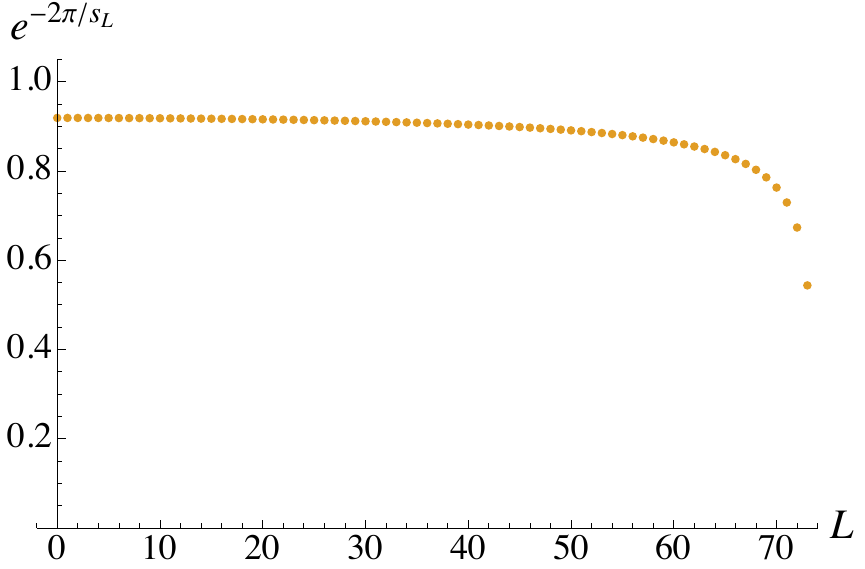}
\caption{\label{fig:scaling}
Discrete scaling factor $e^{-2\pi/s_L}$ for a hydrogen molecular ion obtained from Eq.~(\ref{eq:scaling}) with $d=3$ and $M_1/m=M_2/m=1836$, which ranges from $0.919$ at $L=0$ to $0.544$ at $L=73$ within the allowed angular momentum.}
\end{figure}

In particular, the high-precision spectroscopy of $\H_2^+$ with its ground-state energy being $E_{\H_2^+}=-0.597\,\Eh$ ($\Eh\equiv\hbar^2/ma_0^2=27.21\,\eV$) may reveal the Efimovian states as a sequence of resonances at $E_N=-0.125\,\Eh-(\hbar^2\kappa_L^{*2}/2M)\,e^{-2\pi N/s_L}$ for each $L\leq73$ with the discrete scaling factor in Fig.~\ref{fig:scaling}, which accumulate toward the second dissociation threshold corresponding to $(\H)_{n=2}+\H^+$.
Similarly, multiple sequences of Efimovian resonances accumulating toward every dissociation threshold at $\E_{n\geq3}=-\Eh/2n^2$ corresponding to $(\H)_{n\geq3}+\H^+$ are also expected.
We plan to study their observability in detail as future work.
Furthermore, it is interesting to point out that a hydrogen molecular ion has an extremely shallow $s$-wave bound state, which makes the scattering length between a hydrogen atom and a proton as large as $750\,a_0$~\cite{Carbonell:2003}.
Therefore, the Efimov effect of two hydrogen atoms and one proton may, in principle, be discussed~\cite{Macek:2007}, so that the hydrogen molecular ion constitutes a unique system possibly linked to both Efimov and Efimovian physics.

We also showed that the same system but with a logarithmic Coulomb potential forms an infinite sequence of loosely bound states obeying discrete scale invariance, which are now realized as genuine bound states accumulating toward the first dissociation threshold from below.
Provided that such Efimovian states for a large mass ratio survive even down to equal masses, they may be observed with quantum vortices in two-dimensional superfluids~\cite{Gauthier:2019,Johnstone:2019}.
Our findings hopefully pioneer Efimovian physics emergent from long-range potentials of charged particles.

\acknowledgments
This work was supported by JSPS KAKENHI Grants No.\ JP18H05405 and No.\ JP21K03384.

\section*{Appendix: Hydrogenlike atom}
Here we review the bound-state solutions to the hydrogenlike problems,
\begin{align}
\E\,\psi(\r) = \left(-\frac{\hbar^2\nabla_{\!\r}^2}{2m} - \frac{\hbar^2}{ma_0r}\right)\psi(\r),
\end{align}
in all $d=1$, 2, and 3 dimensions~\cite{Landau-Lifshitz,Yang:1991,Loudon:1959}.
By separating the radial and angular variables as
\begin{align}\label{eq:separation}
\psi(\r) = X(r) \times
\begin{dcases}
\,Y_\l^{m_\l}(\theta,\phi) & (d=3,\ \l\in\N_0,\ |m_\l|\leq\l), \\
\ \frac{e^{i\l\phi}}{\sqrt{2\pi}} & (d=2,\ \l\in\Z), \\
\,\frac{[\sgn(x)]^\l}{\sqrt2} & (d=1,\ \l=0,1),
\end{dcases}
\end{align}
the radial wave function solves
\begin{align}
\kappa^2X(r) = \left[\frac{d^2}{dr^2} + \frac{d-1}{r}\frac{d}{dr}
- \frac{\l(\l+d-2)}{r^2} + \frac2{a_0r}\right]X(r),
\end{align}
where $\kappa\equiv\sqrt{-2m\E/\hbar^2}$.
Then, by substituting $X(r)=\rho^{|\l|}e^{-\rho/2}Z(\rho)$ with $\rho\equiv2\kappa r$, the radial Schr\"odinger equation can be brought into the Laguerre differential equation in the form of
\begin{align}
& \left[\rho\frac{d^2}{d\rho^2} + (2|\l|+d-1-\rho)\frac{d}{d\rho}\right. \notag\\
&\quad\left.{} - \left(|\l|+\frac{d-1}{2}\right) + \frac1{\kappa a_0}\right]Z(\rho) = 0.
\end{align}

In order for the bound-state wave function to be convergent at $r\to\infty$,
\begin{align}
\frac1{\kappa a_0} - \left(|\l|+\frac{d-1}{2}\right) = \nu \in \N_0
\end{align}
must be a non-negative integer~\cite{Laguerre}, so that the binding energy is found to be
\begin{align}\label{eq:hydrogen}
\E_n = -\left(n-\frac{3-d}{2}\right)^{-2}\frac{\hbar^2}{2ma_0^2},
\end{align}
where the principal quantum number is introduced via $n\equiv \nu+|\l|+1$.
The corresponding wave function reads
\begin{align}
Z_{n\l}(\rho) = \sqrt{\frac{(2\kappa)^d(n-|\l|-1)!}{(2n+d-3)(n+|\l|+d-3)!}}\,
L_{n-|\l|-1}^{2|\l|+d-2}(\rho),
\end{align}
which is normalized as~\cite{Laguerre}
\begin{align}
\int_0^\infty\!dr\,r^{d-1}X_{n\l}(r)X_{n'\l}(r) = \delta_{nn'}.
\end{align}
The ground state at $n=1$ takes $\l=0$ only and is nondegenerate, whereas the excited states at $n\geq2$ are $n^2$-fold degenerate for $d=3$, $2n{-}1$-fold degenerate for $d=2$, and twofold degenerate for $d=1$.
We note that the ground-state energy for $d=1$ is divergent because the Coulomb potential is singular at the origin, which is made finite by removing the singularity, for example, with the replacement of $k_eq^2/r\to k_eq^2/\sqrt{r^2+\delta^2}$~\cite{Loudon:1959,Loudon:2016}.
Although the twofold degeneracy at $n\geq2$ is lifted by the regularized Coulomb potential, it is to be restored in the limit of $\delta\to0$.

\end{document}